\documentclass[preprint,preprintnumbers,amsmath,amssymb]{revtex4}

\usepackage{epsfig}
\usepackage{tikz}
\usepackage{graphicx} 
\usepackage{dcolumn} 
\usepackage{bm} 
\usepackage{url}
\usepackage{xcolor}

\begin{document}

\title{Assessing high-order effects in feature importance via predictability decomposition}

\author{Marlis Ontivero-Ortega$^{1}$, Luca Faes$^{2}$, Jesus M Cortes$^{3}$, Daniele Marinazzo$^{4}$, and Sebastiano Stramaglia$^{1}$}

\affiliation{$^1$ Dipartimento Interateneo di Fisica, Universit\`a degli Studi di Bari Aldo Moro, and INFN, Sezione di Bari, 70126 Bari, Italy\\}

\affiliation{$^2$ Dipartimento di Ingegneria, Universit\`a di Palermo, 90128, Palermo, Italy\\
Faculty of Technical Sciences, University of Novi Sad, Serbia\\}

\affiliation{$^3$ Biocruces-Bizkaia Health Research Institute, Barakaldo, Spain\\
Biomedical Research Doctorate Program, University of the Basque Country, Leioa, Spain\\
Department of Cell Biology and Histology, University of the Basque Country, Leioa, Spain\\
IKERBASQUE Basque Foundation for Science, Bilbao, Spain\\}

\affiliation{$^4$  Department of Data Analysis, Ghent University, Ghent, Belgium\\}

\date{\today}

\begin{abstract} 
Building on recent advances in describing redundancy and synergy in multivariate interactions among random variables, we propose a novel approach to quantify cooperative effects in feature importance, a key technique in explainable artificial intelligence. Specifically, we introduce an adaptive version of the widely used metric Leave One Covariate Out (LOCO), designed to disentangle high-order effects involving a particular input feature in regression problems. LOCO measures the reduction in prediction error when the feature of interest is added to the set of features used in regression. Unlike the standard approach that computes LOCO using all available features, our method identifies the subsets of features that maximize and minimize LOCO. This results in a decomposition of LOCO into a two-body component and higher-order components (redundant and synergistic), while also identifying the features that contribute to these high-order effects in conjunction with the driving feature. We demonstrate the effectiveness of the proposed method in a benchmark data-set related to wine quality and to proton/pion discrimination using simulated detector measurements generated by GEANT.

\end{abstract}\maketitle

\maketitle
In explainable artificial intelligence (XAI) \cite{xai}, global feature importance methods \cite{fi} assess how much each individual feature contributes to predicting the target variable, providing a single importance score for each feature. Among the most commonly used metrics for feature importance is Leave One Covariate Out (LOCO) \cite{loco}, which measures the reduction in prediction error when the feature of interest is added to the set of features used in regression.

However, accurately predicting the target variable often requires accounting for interactions among multiple features. Furthermore, input features may exhibit strong correlations with one another, and summarizing cooperative effects in a single score for each feature can lead to misinterpretations. To address this, we propose a novel decomposition of feature importance scores that separates three components: the unique, pure (two-body) influence of a feature on the target, as well as the contributions arising from synergistic and redundant interactions with other features.

Our approach exploits recent results which generalize the traditional dyadic description of networks of variables to the higher-order setting \cite{battiston,rosas}, with an increasing attention devoted to the emergent properties of complex systems, manifesting through high-order behaviors sought in observed data. A key framework in this literature is the partial information decomposition (PID) \cite{pid} and its subsequent developments \cite{lizier}, which utilize information-theoretic tools to reveal high-order dependencies among groups of three or more  variables and describe their synergistic or redundant nature. Within this framework, redundancy refers to information retrievable from multiple sources, while synergy refers to information retrievable from the whole system that cannot be observed in its individual parts. Notably, in some papers the use of predictability (from linear modeling of data) has also been proposed to highlight high-order effects \cite{redund,faespred,unnorm}: using reduction in variance instead of entropy, the effect due to a group of independent variables is equal to the sum of effects from the individual variables (assuming linearity), see the related discussion in \cite{barrett}. 

The core concept of the proposed approach, named Hi-Fi (High-order Interactions for Feature Importance), is that the gain in predictability from including a specific variable can be quantified either through conditional mutual information (as a reduction in entropy) or LOCO (as a reduction in error variance). These two measures capture the same underlying effect, with the former grounded in information theory and the latter in regression. Consequently, substituting conditional mutual information with LOCO provides a novel framework for analyzing synergy and redundancy within the context of regression theory.

Recently, a decomposition of transfer entropy \cite{schreiber}—a primary method for evaluating the {\it reduction in surprise} between random processes—into unique, redundant, and synergistic components has been proposed \cite{disentangling}. This decomposition allows for the quantification of the relative importance of high-order effects compared to pure two-body effects in information transfer between two processes, while also identifying the processes that contribute to these high-order effects alongside the driver.

Adapting this framework by replacing conditional mutual information with LOCO enables a decomposition of LOCO into unique, redundant, and synergistic components, achieving the desired decomposition of feature importance. Substituting entropy with variance offers an alternative definition of synergy and redundancy, rooted in regression. This approach avoids distortions that arise when comparing combined information from two independent sources to the sum of information from each source individually, which result from the concavity of the logarithmic function. However, this comes at the cost of sacrificing the formal rigor of information theory \cite{barrett}.


In the following, we provide a detailed description of the proposed methodology.

Consider $n$ stochastic variables ${\bf Z}=\{z_\alpha \}_{\alpha=1,\ldots,n}$, a driver variable $X$, and a target variable $Y$. Let $\epsilon \left(Y|X,{\bf Z}\right)$ represent the mean squared prediction error \cite{nota} of $Y$ based on all input variables, $X$ and ${\bf Z}$ (applicable to linear or non-linear regression models). Now, consider $\epsilon \left(Y |{\bf Z}\right)$, the prediction error of $Y$ based solely on the variables in ${\bf Z}$. The difference in errors under these two conditions is given by:

\begin{equation}\label{loco} L_{\bf Z}(X\to Y) =\epsilon \left(Y |{\bf Z}\right)-\epsilon \left(Y|X,{\bf Z}\right) \end{equation}

This expression defines the well-known measure of feature importance, {\it Leave One Covariate Out} (LOCO) \cite{loco}, which is non-negative for a broad class of predictors \cite{invariance}. Equation (\ref{loco}) quantifies the reduction in predictive power when feature $X$ is excluded from the regression model. Consequently, $L_{\bf Z}(X\to Y)$ serves a similar role in predictability as the conditional mutual information $I(X;Y|{\bf Z})$ does in entropy reduction. Replacing $I(X;Y|{\bf Z})$ with $L_{\bf Z}(X\to Y)$ in information-theoretic frameworks for high-order dependencies introduces an alternative definition of synergy, interpreted in terms of predictability and regression.

Using the same notation, the reduction in error variance considering only the driver $X$ (referred to as the {\it pairwise} predictive power or explained variance) is denoted as $L_{\bf \emptyset}(X\to Y) = \sigma^2_Y - \epsilon \left(Y|X\right)$. The relationship between $L_{\bf \emptyset}(X\to Y)$ and $L_{\bf Z}(X\to Y)$ can indicate redundancy or synergy: $L_{\bf \emptyset}(X\to Y) > L_{\bf Z}(X\to Y)$ implies redundancy among $X$, ${\bf Z}$, and $Y$, whereas $L_{\bf \emptyset}(X\to Y) < L_{\bf Z}(X\to Y)$ suggests synergy.

As discussed in \cite{disentangling}, $L_{\bf Z}(X\to Y)$ underestimates the importance of $X$ when $X$ exhibits redundancy with some variables in ${\bf Z}$. Conversely, considering only the driver $X$ (i.e., $L_{\bf \emptyset}(X\to Y)$) overlooks synergies between $X$ and ${\bf Z}$. In other words, it neglects suppressor variables \cite{suppressor} in ${\bf Z}$ that could enhance the predictability of $Y$ based on $X$. 

Considering now just a subset ${\bf z}$ of all the variables in ${\bf Z}$, it is intuitive that searching for  ${\bf z_{min}}$ minimizing $L_{\bf z}(X\to Y)$  captures the amount of redundancy $R$ that the rest of variables share with the pair $X-Y$, i.e. we may define  $R=L_{\bf \emptyset}(X\to Y)-L_{\bf z_{min}}(X\to Y)$. The unique predictive power $U$, i.e. the pure two-body influence of $X$ on $Y$, will be given by $U=L_{\bf z_{min}}(X\to Y)$. On the other hand, searching for ${\bf z_{max}}$ maximizing $L_{\bf z}(X\to Y)$, leads to the amount of synergy $S$ that the {\it remaining variables} provide in terms of the increase of predictability:  $S=L_{\bf z_{max}}(X\to Y) - L_{\bf \bf \emptyset}(X\to Y)$. 

It follows that:
\begin{equation}
L_{\bf z_{max}}(X\to Y)= S+R+U,\\
\label{eq:decomposition}
\end{equation}
and the maximal predictive power of $X$ to $Y$ can be decomposed into the sum of a unique contribution (U), representing a pure two-body effect, synergistic (S) and redundant (R) contributions that describe cooperative effects on $Y$, due to $X$ and ${\bf Z}$ variables. 

Conducting an exhaustive search for subsets ${\bf z_{min}}$ and ${\bf z_{max}}$ becomes unfeasible for large $n$. Here, we suggest  employing a greedy  strategy, whereby we perform a search over all the ${\bf z}$ variables for the first variable to be tentatively used. Subsequently, variables are added one by one, to the previously selected ones, to construct the set of ${\bf z}$ variables that either maximize or minimize the $L_{\bf z}$. 
The criterion for terminating the greedy search, minimizing (maximizing) the $L_{\bf z}$, is to stop when the corresponding decrease (increase) is compatible with a random effect.
Therefore one can estimate the probability that the increase in $L_{\bf z}$ is lower (higher) than the one corresponding to the inclusion of a variable sharing the individual statistical properties of the selected one but being otherwise uncoupled from $X$ and $Y$ (surrogates of the selected $z$ variable are obtained by permutation). If this probability is below a given threshold (corrected for multiple comparisons), the selected variable is thus added to the multiplet. 

As a toy problem to demonstrate the proposed approach, we consider three zero-mean, unit-variance Gaussian variables, $a_1$, $a_2$, and $a_3$, with correlations $\langle a_1 a_2\rangle = 0.5$, $\langle a_1 a_3\rangle = 0.3$, and $\langle a_2 a_3\rangle = -0.5$. In this setup, $a_3$ acts as a suppressor for $a_2$ \cite{sup1} (and vice versa, $a_2$ is a suppressor for $a_3$), forming a synergistic triplet. Additionally, we define $b_1$, $b_2$, and $b_3$ as another set of variables with correlations $\langle b_1 b_2\rangle = 0.5$, $\langle b_1 b_3\rangle = 0.3$, and $\langle b_2 b_3\rangle = 0.5$. These variables constitute a redundant triplet since they are all positively correlated. Let $c$ be an independent source, and $\{d_1, d_2\}$ represent two independent sources.

The target variable $Y$ is defined as $$Y=a_1+b_1+c+d_1 d_2 +\eta,$$
where $\eta$ is a Gaussian noise term with a standard deviation of 0.05, and $\{d_1, d_2\}$ contribute synergistically to $Y$. The input variables are denoted as: $X_1=a_2$, $X_2=a_3$, $X_3=b_2$, $X_4=b_3$, $X_5=c$, $X_6=d_1$, and $X_7=d_2$.

Given the structure of this example, we use kernel regression with an inhomogeneous polynomial kernel of degree 2 as the predictive model. This corresponds to linear regression augmented with features consisting of all monomials of the input variables with degrees less than or equal to two \cite{shawe}.

\begin{figure}[!ht] \centering \includegraphics[width=.95\textwidth]{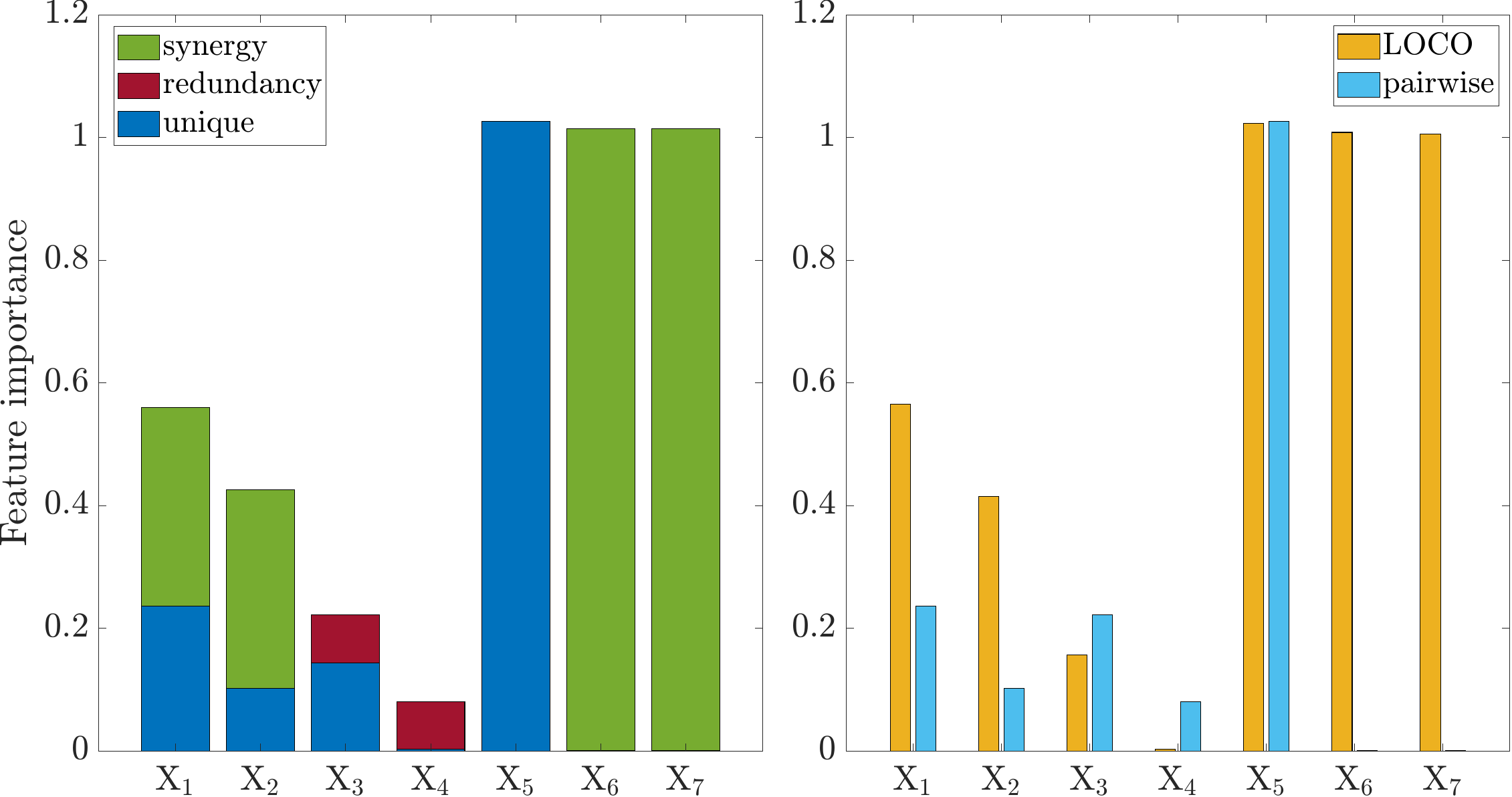} \caption{Left: The predictability decomposition by Hi-Fi, for the toy problem, is depicted for the seven driving variables. Right: The LOCO and the pairwise index are depicted for the seven variables.}
\label{fig:decomp_toy}\end{figure}

\begin{figure}[!ht] \centering \includegraphics[width=.85\textwidth]{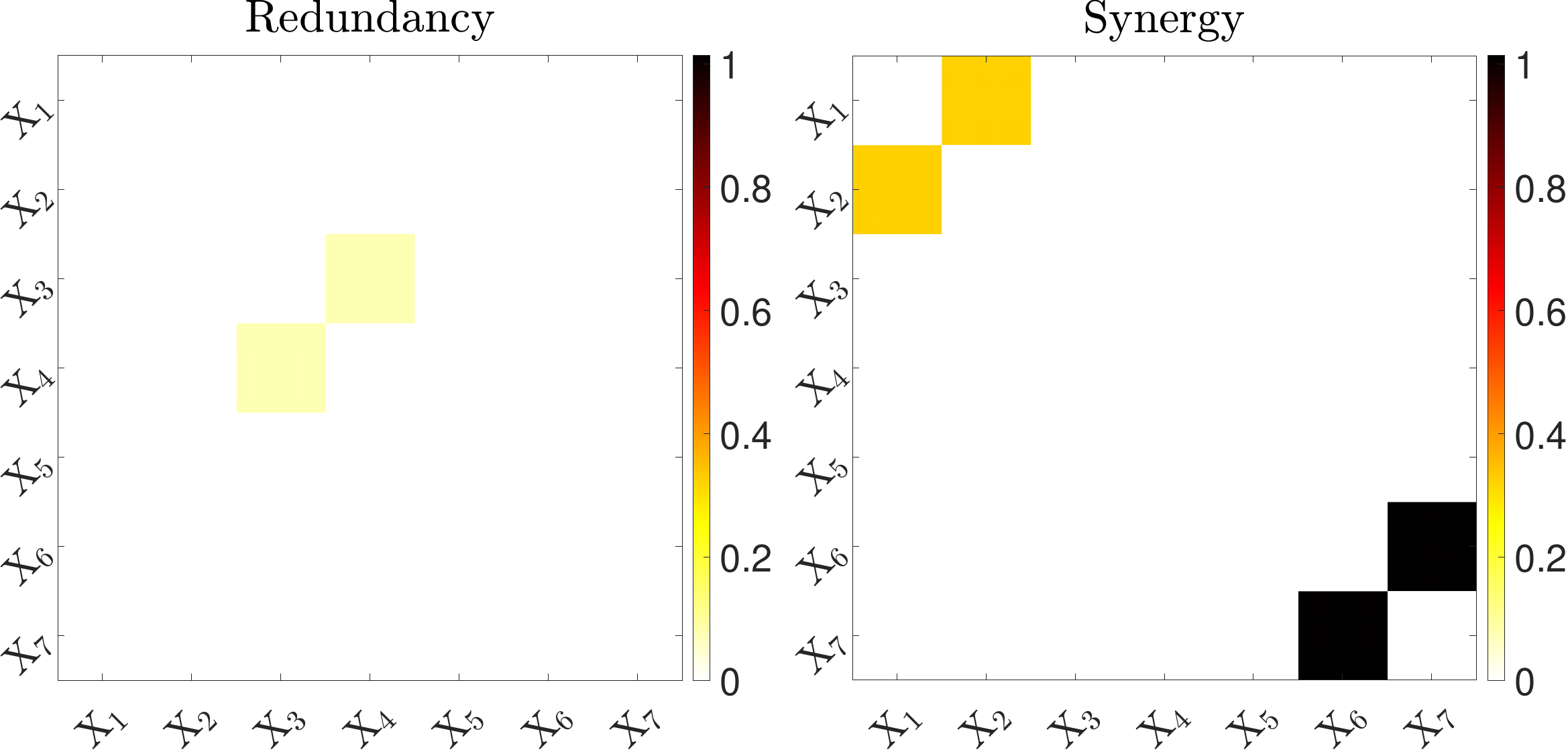} \caption{For each driving variable in the toy example (each row), the color value represents the decrement (increment) of LOCO due to the inclusion of each other variable in the redundant (synergistic) multiplet.}\label{fig:multiplets_wine} \end{figure}

In Figure \ref{fig:decomp_toy}, we present the results of Hi-Fi for the seven driving variables, along with the LOCO and the pairwise predictive power for each variable. The proposed approach effectively highlights both the importance and the nature of each driver's contribution to predicting the target. Specifically, $X_1$ and $X_2$ are identified as synergistic, while $X_3$ and $X_4$ are recognized as redundant. Notably, all four variables also provide unique contributions. Meanwhile, $X_5$ contributes only unique information to the target, and $X_6$ and $X_7$ are correctly classified as providing purely synergistic information. As illustrated in the right panel, LOCO underestimates the importance of the redundant variables $X_3$ and $X_4$, whereas the pairwise index fails to capture the synergistic contributions of $X_1$, $X_2$, $X_6$, and $X_7$.

It is worth emphasizing the difference between the two pairs of synergistic variables, ${X_1, X_2}$ and ${X_6, X_7}$. The target $Y$ is linearly dependent on $X_1$ and $X_2$, making their synergy a result of dependency. If $X_1$ and $X_2$ were independent, their synergy, as determined by the proposed method, would be zero. In contrast, $X_6$ and $X_7$ are independent, and their synergy arises from the nonlinear interaction term $d_1 d_2$, which structurally contributes to $Y$.


The proposed framework was evaluated on two different publicly available data-sets. In both cases, all predictor variables were z-scored prior to proceeding with the decomposition.
 As a first example, we used the Wine Quality data set (n = 6497) obtained from the UCL Machine Learning Repository \cite{wine_dataset}. The goal is to predict wine quality scores and
we employed a simple linear regression model as a function to predict the response variable and estimate the error variance. 

\begin{figure}[!ht] \centering
\includegraphics[width=.95\textwidth]{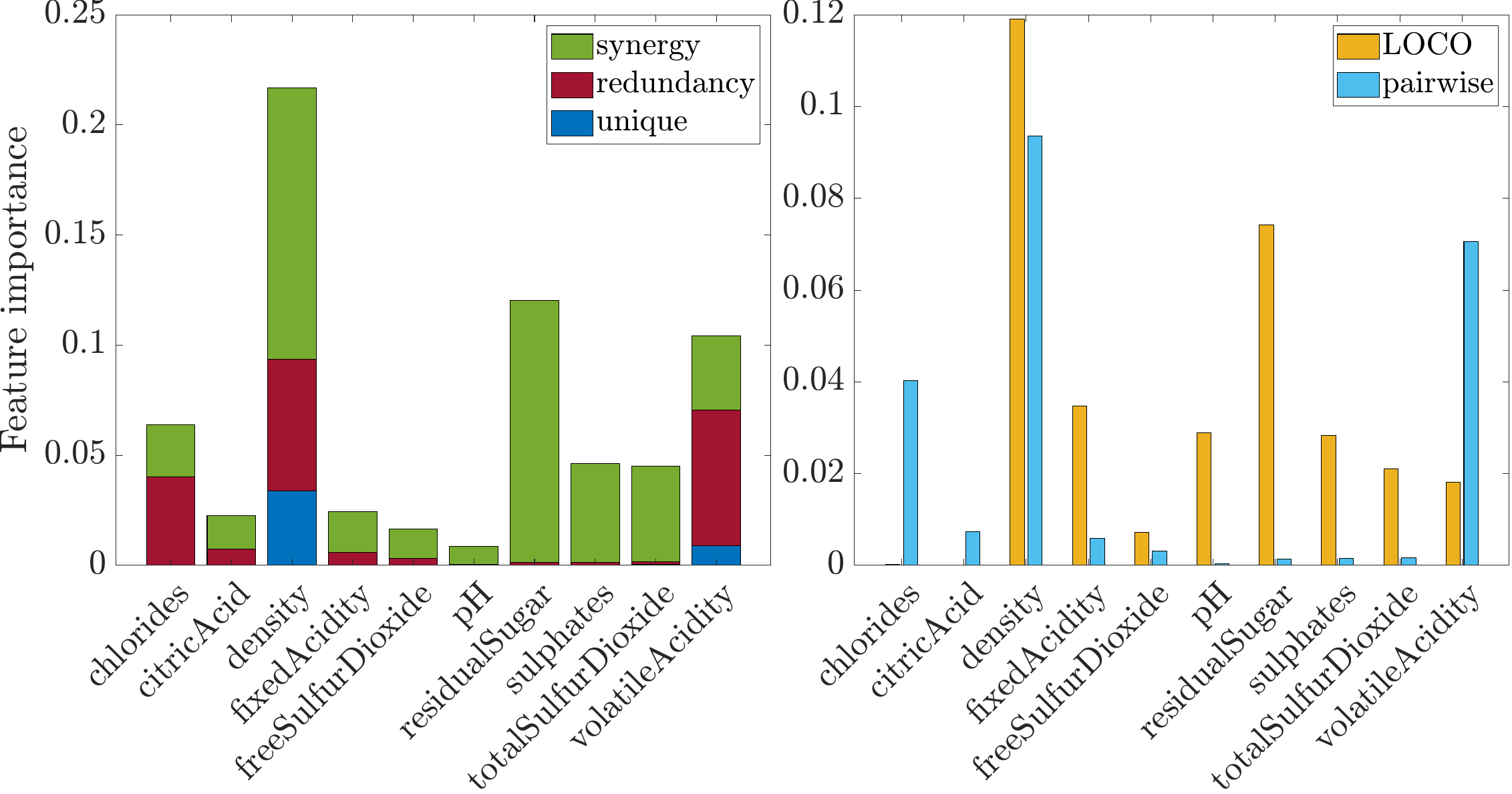}
\caption{Left: The predictability decomposition by Hi-Fi, for the particle discrimination problem, is depicted for the ten driving variables. Right: The LOCO and the pairwise index are depicted for the ten variables.}
\label{fig:decomp_wine}
\end{figure}

The most important feature is the density, which shows the largest unique contribution to the LOCO and is also the most as synergistic variable together with  residual sugar. Density, chlorides and volatile acidity are the most redundant variables. On the other hand, the LOCO index fails to highlight redundant effects, indeed the importance of density, chlorides and volatile acidity are underestimated; moreover the pairwise index does not put in evidence the synergistic role of residual sugar.

\begin{figure}[!ht] \centering \includegraphics[width=.85\textwidth]{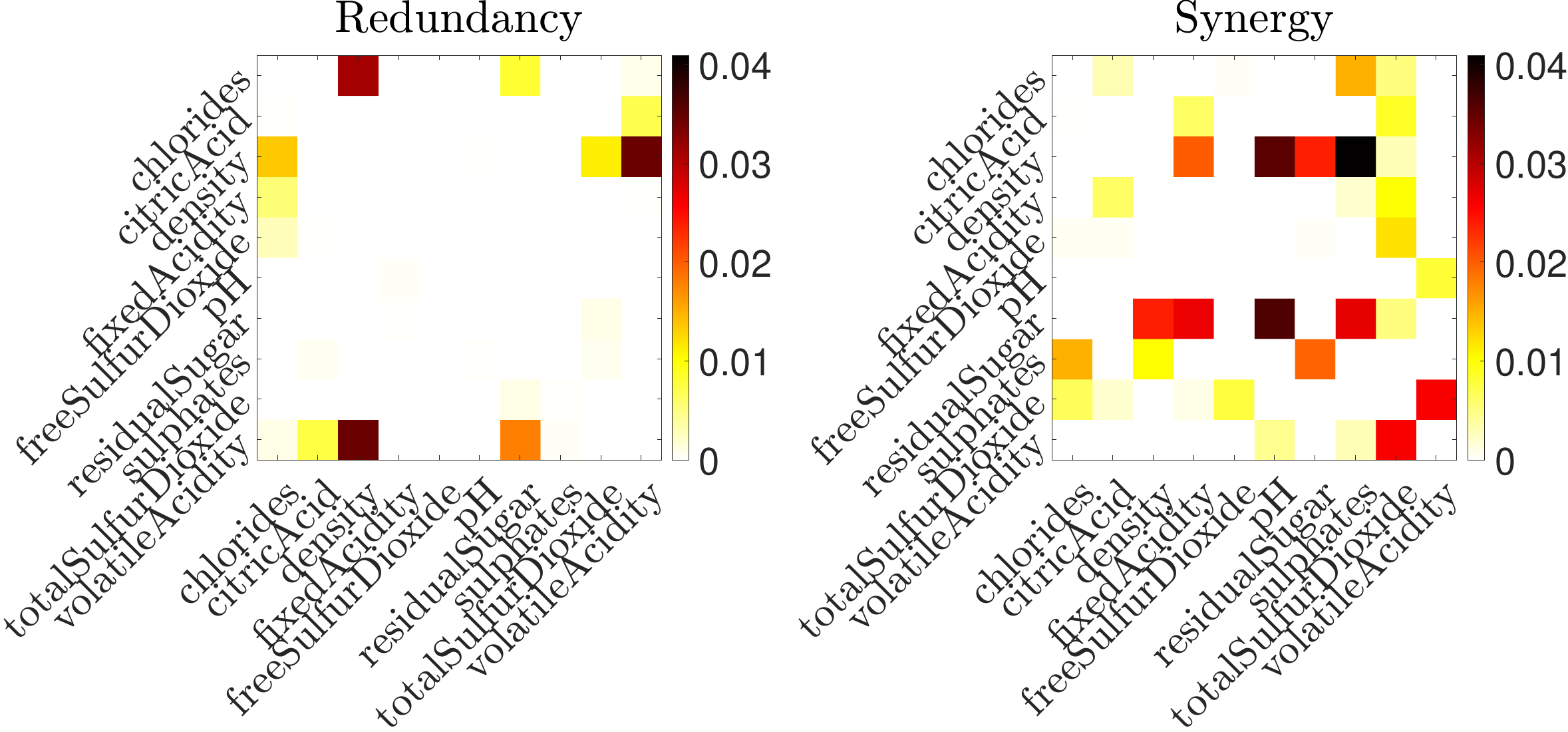} \caption{For each driving variable in the wine quality problem (each row), the color value represents the decrement (increment) of LOCO due to the inclusion of each other variable in the redundant (synergistic) multiplet.}\label{fig:multiplets_wine} \end{figure}
 
We now consider a dataset derived from a GEANT \cite{GEANT} simulation of electron-proton inelastic scattering, measured by a particle detector system and available in \cite{geant}. This dataset includes a very large number of samples of four different particles, each characterized by six detector signals. For this analysis, we focus exclusively on pion and proton samples (n=4,752,682) to predict a binary target variable, where y=1 represents protons and y=0 represents pions. As in the previous toy example, we use the hypothesis space induced by the inhomogeneous polynomial kernel of degree 2 (similar outputs are obtained using linear regression). The input features include the velocity $\beta$, the momentum $p$, the scattering angle $\theta$, the number of emitted photoelectrons $nphe$, the response from an inner detector $ein$, and the response from an outer detector $eout$. 

\begin{figure}[!ht] \centering
\includegraphics[width=.95\textwidth]{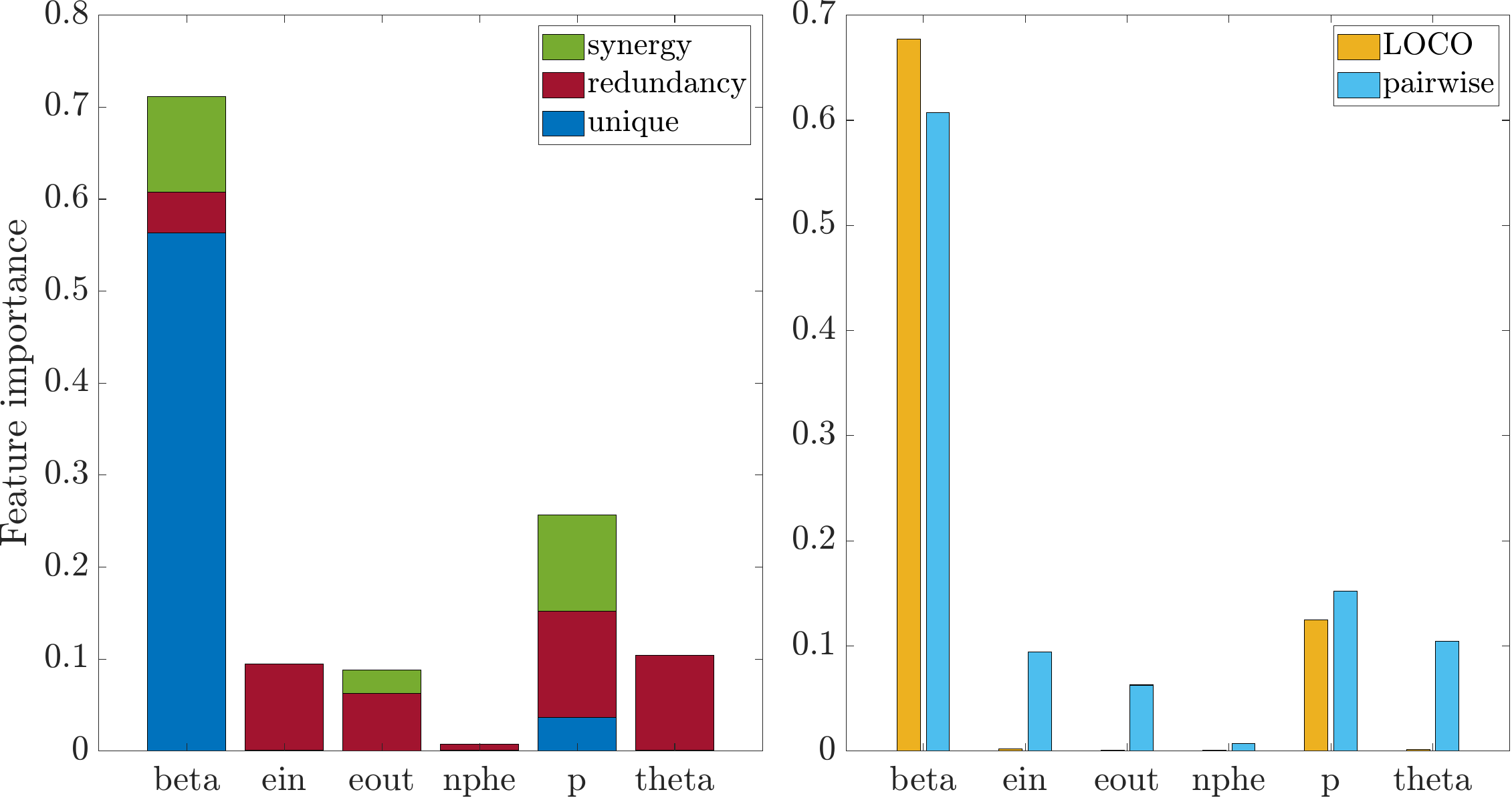}
\caption{Left: The predictability decomposition by Hi-Fi, for the particle discrimination problem, is depicted for the six driving variables. Right: The LOCO and the pairwise index are depicted for the six variables.}
\label{fig:decomp_geant}
\end{figure}

In Figure \ref{fig:decomp_geant}, we depict the decomposition by Hi-Fi along with LOCO and the pairwise index for the six variables. According to the pairwise index, the features with the highest predictive power are, in order, $\beta$, $p$, and $\theta$, while $ein$ and $eout$ also demonstrate predictive power. The feature $nphe$ shows negligible predictive power. LOCO, on the other hand, identifies $\beta$ and $p$ as the most significant features. Using the Hi-Fi approach, the most discriminating feature is $\beta$, which provides a large unique contribution, synergistic effects when combined with $p$, and redundant effects with $\theta$.

\begin{figure}[!ht] \centering \includegraphics[width=.85\textwidth]{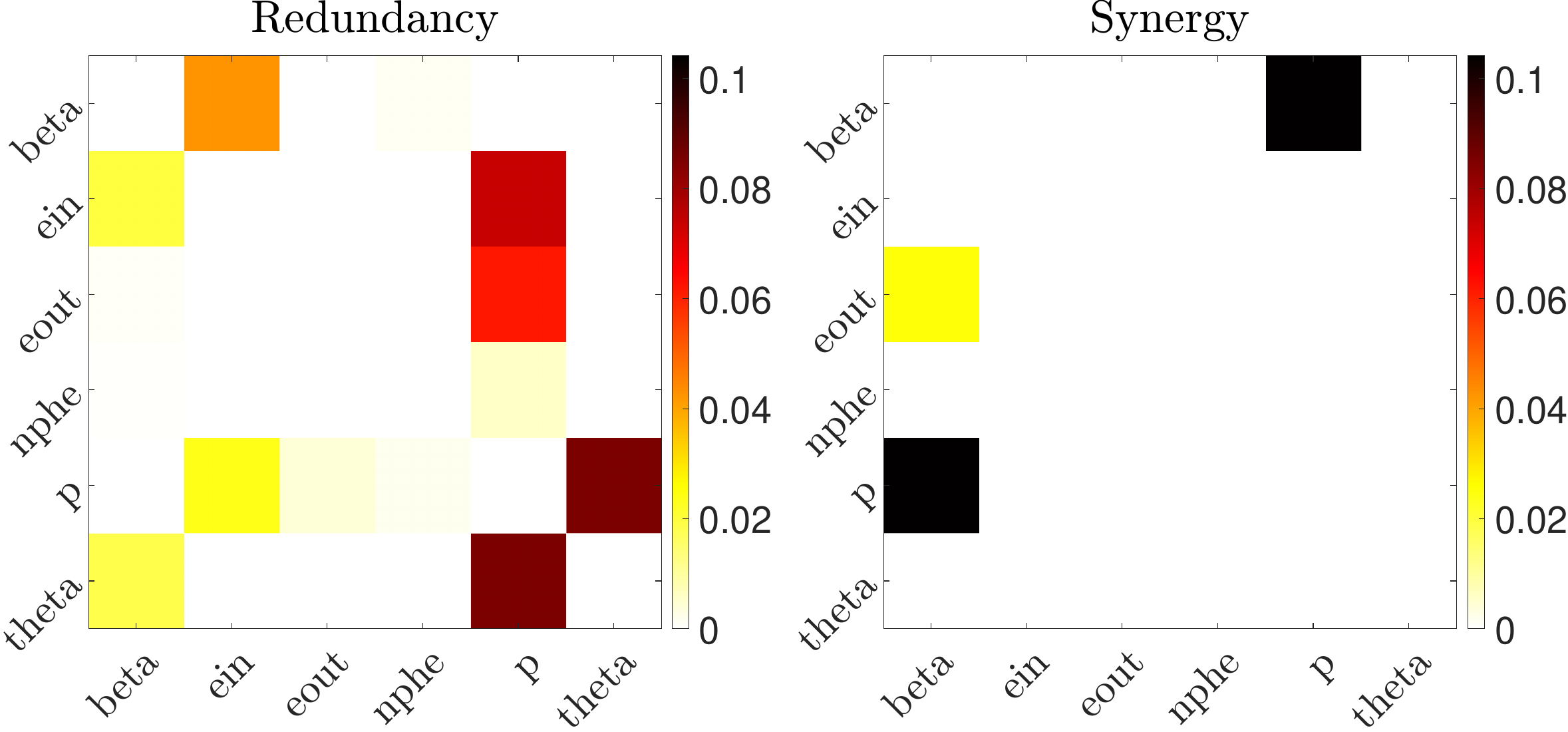} \caption{For each driving variable in the particle discrimination problem (each row), the color value represents the decrement (increment) of LOCO due to the inclusion of each other variable in the redundant (synergistic) multiplet.} \label{fig:multiplets_geant} \end{figure}

The momentum $p$ demonstrates low importance when considered alone (small $U$), but it becomes relevant through synergy in cooperation with $\beta$ and redundancy with $\theta$ and $ein$. The feature $eout$ exhibits synergy with $\beta$, which can be attributed to the fact that protons tend to release energy in the inner detectors, whereas pions are more likely to release energy in the outer detector. The redundancy of $ein$ and $eout$ with $p$ and $\beta$ is consistent with the Bethe-Bloch formula, which describes the energy loss of charged particles as they pass through matter. Additionally, $\theta$ is redundant with $\beta$ and $p$, indicating that once $\beta$ and $p$ are known, the kinematics of the scattering does not provide further information about the particle. The feature $nphe$ is found to be almost irrelevant. These results are further described in Figure \ref{fig:multiplets_geant}, where the subsets of variables resulting in a modification of LOCO are depicted for each driving variable.

Compared to LOCO, our approach reveals the significance of $ein$, $eout$, and $\theta$ despite their redundancy with $\beta$ and $p$. Furthermore, it highlights the synergistic interaction between $\beta$ and $p$ in enhancing the accuracy of particle discrimination. The combined knowledge of $\beta$ and $p$ enables inference of the particle's mass, suggesting that the synergistic information identified by our method is related to the particle mass.


Interestingly, a recent decomposition of LOCO has been proposed to disentangle interactions and dependencies in feature attribution (DIP) \cite{konig}. In this framework, LOCO is expressed as the sum of each feature’s standalone contribution, the contributions due to interactions, and those arising from main effect dependencies. However, 
the dependency contributions aggregate both redundant and synergistic effects, while interaction-based synergistic contributions are treated separately.

In contrast, our decomposition focuses on disentangling redundancy and synergistic effects explicitly. In our framework, synergy encompasses both dependency effects and interaction effects. As such, the proposed approach and the decomposition described in \cite{konig} can be viewed as complementary. Furthermore, unlike the approach in \cite{konig}, our method identifies the specific variables that contribute to redundant and synergistic effects alongside the feature under consideration.

It is worth noting that the proposed decomposition, just like vanilla LOCO, is influenced by the choice of the hypothesis space for regression: performing the analysis with multiple models may be useful to assess the robustness of the conclusions w.r.t. the choice of the model. We like to stress that the exhaustive evaluation is feasible on the data-sets analyzed in this work, as the number of variables is not huge: we obtained the same results as those from the greedy search thus confirming the validity of the greedy approach.

In summary, we have addressed feature importance for regression, a critical tool in XAI, by introducing a novel definition of importance that incorporates the cooperative behavior among features. The total importance is decomposed into a unique contribution from the feature itself and the redundant and synergistic contributions that quantify its cooperation with other features.


\clearpage

\section*{Data and code availability}
The code to simulate and analyze data is available at \url{https://github.com/codesCN/Hi-Fi}

\begin{acknowledgments}
We thank dr. Nicola Colonna (Istituto Nazionale Fisica Nucleare, Sezione di Bari, Italia) for useful discussions about GEANT.

LF, MOO and SS were supported by the project “HONEST - High-Order Dynamical Networks in Computational Neuroscience and Physiology: an Information-Theoretic Framework”, Italian Ministry of University and Research (funded by MUR, PRIN 2022, code 2022YMHNPY, CUP: B53D23003020006).
 SS was supported by the project “Higher-order complex systems modeling for personalized medicine”, Italian Ministry of University and Research (funded by MUR, PRIN 2022-PNRR, code P2022JAYMH, CUP: H53D23009130001). \end{acknowledgments}


\begin{thebibliography}{99}
\bibitem{nota} In general it is preferable to estimate the mean squared error by cross validation, however when the number of samples is sufficiently high, like in the examples reported in this paper, evaluation on the training data  may be used as well.
\end{thebibliography}
\end{document}